\documentclass[sigconf]{acmart}

\AtBeginDocument{%
  }

\setcopyright{none}
\acmYear{2022}

\acmConference[ESEM'22]{16th ACM/IEEE International Symposium on Empirical Software Engineering and Measurement}{September 18--23, 2022}{Helsinki, Finland}

\PassOptionsToPackage{table}{xcolor}

\usepackage[capitalise]{cleveref}

\usepackage{paralist}

\usepackage[inline]{enumitem}  
\newlist{researchquestions}{enumerate}{2}

\setlist[researchquestions,1]{label=RQ\arabic*, font=\itshape\bfseries, align=left, leftmargin=*}
\setlist[researchquestions,2]{label*=.\alph*, font=\itshape\bfseries, align=left, leftmargin=.5em}

\usepackage[explicit]{titlesec}
\titleformat{\paragraph}[runin]{\normalfont\normalsize\bfseries}{\theparagraph}{1em}{#1.~}
\titlespacing*{\paragraph}{0pt}{0ex plus 0.2ex minus .2ex}{1ex}

\usepackage{amsthm}

\theoremstyle{remark}
\newtheorem{example}{Example}

\usepackage{caption}
\usepackage{subcaption}

\usepackage{xspace} 

\newcommand{\eg}{e.g.\xspace}

\newcommand{\ie}{i.e.\xspace}

\newcommand{\wrt}{w.r.t.\xspace}

\newcommand{\DM}{\ac{dm}\xspace}  
\newcommand{\DMs}{\acp{dm}\xspace} 
\newcommand{\BD}{\ac{bd}\xspace} 
\newcommand{\TS}{\ac{ts}\xspace} %
\newcommand{\TSs}{\acp{ts}\xspace} %
\newcommand{\RQ}{\ac{rq}\xspace} %

\newcommand{\sbst}{\texttt{SBST}\xspace}
\newcommand{\sbstcomp}{\texttt{SBST'2022 Tool Competition}\xspace}

\newcommand{\ambiegen}{\texttt{AmbieGen}\xspace}
\newcommand{\wogan}{\texttt{WOGAN}\xspace}
\newcommand{\frenetic}{\texttt{Frenetic}\xspace}
\newcommand{\beamng}{\texttt{BeamNG.AI}\xspace}
\newcommand{\davetwo}{\texttt{Dave2}\xspace}

\newcommand{\road}[1]{\textbf{\texttt{#1}}\xspace}  
\newcommand{\resq}[1]{\textbf{\textit{RQ#1}}\xspace}  

\newcommand{\numDMs}{\ensuremath{47}\xspace}
\newcommand{\numroads}{\ensuremath{97{,}000}\xspace}

\usepackage[acronyms,shortcuts,section=section,style=super,nopostdot,nogroupskip,nomain,nonumberlist]{glossaries}
\glsdisablehyper

\newacronym{ads}    {ADS}   {automated driving system}
\newacronym{adas}   {ADAS}  {advanced driver-assistance system}
\newacronym{aads}   {AADS}  {automated and autonomous driving systems}
\newacronym{av}     {AV}    {autonomous vehicle}
\newacronym{bd}     {BD}    {behavioural diversity}

\newacronym{cas}    {CAS}   {collision avoidance system}
\newacronym{da}     {DA}    {driving agent}
\newacronym{dm}     {DM}    {diversity measure}

\newacronym{js}     {JS}    {Jaccard similarity index}
\newacronym{rq}     {RQ}    {research question}

\newacronym{sbt}    {SBT}   {search-based testing}
\newacronym{sbse}   {SBSE}  {search-based software engineering}
\newacronym{sut}    {SUT}   {system under test}
\newacronym{ts}    {TS}   {test suite}

\usepackage{tikz}

\usetikzlibrary{positioning}
\usetikzlibrary{calc} 
\usetikzlibrary{arrows}
\usetikzlibrary{arrows.meta}
\usetikzlibrary{shapes}
\usetikzlibrary{arrows.meta}
\usetikzlibrary{patterns}
\usetikzlibrary{backgrounds}

\pgfmathsetmacro{\markingwidth}{.5pt}

\tikzset{
    roadmarking/.style={draw=black!80,line width=\markingwidth},
    centreline/.style={roadmarking,dashed,dash pattern=on 3pt off 2pt},
    basevehicle/.style={rectangle,minimum height=0.5cm,minimum width=0.25cm,inner sep=0pt,},
    ego/.style={basevehicle,execute at begin node={\includegraphics[trim=0.5cm 0 0.5cm 0,clip,width=.25cm]{egoGO.png}}},
    egopath/.style={draw=blue,-latex,thick},
}

\begin{document}

\title[Does Road Diversity Really Matter in Testing Automated Driving Systems? -- A Registered Report]{Does Road Diversity Really Matter in\\Testing Automated Driving Systems?}
\subtitle{A Registered Report}

\author{Stefan Klikovits}
\email{klikovits@nii.ac.jp}
\orcid{0000-0003-4212-7029}
\affiliation{%
  \institution{National Institute of Informatics}
  \city{Tokyo}
  \country{Japan}
}

\author{Vincenzo Riccio}
\email{vincenzo.riccio@usi.ch}
\orcid{0000-0002-6229-8231}
\affiliation{%
  \institution{Universit\`a della Svizzera Italiana (USI)}
  \city{Lugano}
  \country{Switzerland}
}

\author{Ezequiel Castellano}
\email{ecastellano@nii.ac.jp}
\orcid{0000-0002-9604-9997}
\affiliation{%
  \institution{National Institute of Informatics}
  \city{Tokyo}
  \country{Japan}
}

\author{Ahmet Cetinkaya}
\email{ahmet@shibaura-it.ac.jp}
\orcid{0000-0002-1731-8600}
\affiliation{%
  \institution{Shibaura Institute of Technology}
  \city{Tokyo}
  \country{Japan}
}

\author{Alessio Gambi}
\email{alessio.gambi@fh-krems.ac.at}
\orcid{0000-0002-0132-6497}
\affiliation{%
  \institution{IMC University of Applied Science}
  \city{Krems}
  \country{Austria}
}

\author{Paolo Arcaini}
\email{arcaini@nii.ac.jp}
\orcid{0000-0002-6253-4062}
\affiliation{%
  \institution{National Institute of Informatics}
  \city{Tokyo}
  \country{Japan}
}

\renewcommand{\shortauthors}{Klikovits et al.}

\begin{abstract}
{\it Background/Context.}
The use of \acp{ads} in the real world requires rigorous testing to ensure safety. To increase trust, \acp{ads} should be tested on a large set of diverse road scenarios. Literature suggests that if a vehicle is driven along a set of geometrically diverse roads---measured using various \DMs---it will react in a wide range of behaviours, thereby increasing the chances of observing failures (if any), or strengthening the confidence in its safety, if no failures are observed. To the best of our knowledge, however, this assumption has never been tested before, nor have road \DMs been assessed for their properties.\\
{\it Objective/Aim.}
Our goal is to perform an exploratory study on \numDMs currently used and new, potentially promising road \DMs. Specifically, our \aclp{rq} look into the road \DMs themselves, to analyse their properties (\eg \emph{monotonicity}, \emph{computation efficiency}), and to test correlation between \DMs. Furthermore, we look at the use of road \DMs to investigate whether the assumption that diverse test suites of roads expose diverse driving behaviour holds.\\
{\it Method.}
Our empirical analysis relies on a state-of-the-art, open-source \acp{ads} testing infrastructure and uses a data set containing over \numroads individual road geometries and matching simulation data that were collected using two driving agents. By sampling random test suites of various sizes and measuring their roads' geometric diversity, we study road \DMs properties, the correlation between road \DMs, and the correlation between road \DMs and the observed behaviour.
\end{abstract}


\maketitle

\glsresetall
\section{Introduction}

\Acp{ads} are expected to drastically change the transportation industry by reducing the number of accidents, avoiding traffic congestion, and lowering fuel consumption.
Nonetheless, reports of collisions involving \acp{ads}~\cite{DMVCalifornia,nhtsa2022summary} and fatalities~\cite{elonbachman_2020_3685309} emphasise the need for extensive validation and testing of the technology before they can be safely released onto public roads.

Thorough testing explores, \ie \emph{covers}, different aspects of the \ac{sut}'s behaviour, and thus has the potential to find bugs and increase the confidence in the \ac{sut}'s correctness~\cite{Fraser2019}.
Due to the complexity of the \ac{sut}, however, it is generally not possible to directly find test inputs that maximise \BD. Therefore, to increase testing cost-effectiveness, existing research (\eg~\cite{chen2004adaptive,bueno2014diversity,feldt2016test}) proposed to generate test suites that maximise \emph{test diversity}. The underlying assumption of generating diverse tests is that the more diverse the tests within a \TS are, the higher the \ac{sut}'s \BD will be, and thus, more of its functionality will be exercised~\cite{xie2006studying}.
Consequently, diversity-driven approaches such as Novelty search~\cite{lehman2011abandoning} have been applied to generate tests~\cite{feldt2017searching,zohdinasab2021deephyperion}.

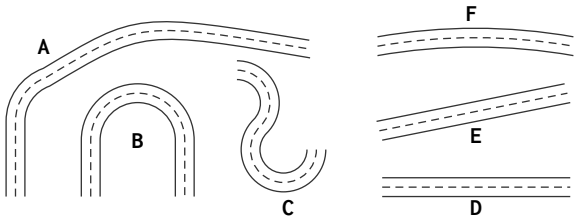
\begin{figure}[t]
\centering
\begin{subfigure}[b]{0.42\textwidth}
\begin{tikzpicture}[scale=0.25, every node/.style={scale=1}]

    \draw[roadmarking] (0,0) -- ++(0,4) arc (180:110:3cm) .. controls ++(5,3) .. ++(14.15,1.5);
    \draw[centreline] (0.5,0) -- ++(0,4) arc (180:110:2.5cm) .. controls ++(5,3) .. ++(14,1.5);
    \draw[roadmarking] (1,0) -- ++(0,4) arc (180:110:2cm) .. controls ++(5,3) .. ++(13.85,1.5);
    \node at (2,8) {\road{A}};

    \tikzset{shift={(4,0)}}
    \draw[roadmarking] (0,0) -- ++(0,3) arc (180:0:3) -- ++(0,-3);
    \draw[centreline] (0.5,0) -- ++(0,3) arc (180:0:2.5) -- ++(0,-3);
    \draw[roadmarking] (1,0) -- ++(0,3) arc (180:0:2) -- ++(0,-3);
    \node at (3,3) {\road{B}};

    \tikzset{shift={(12,0)}}
    \draw[roadmarking] (0,2.5) arc (360:135:1.25) arc (-45:90:2.25);
    \draw[centreline] (0.5,2.5) arc (360:135:1.75) arc (-45:90:1.75);
    \draw[roadmarking] (1,2.5) arc (360:135:2.25) arc (-45:90:1.25);
    \node at (-1,-0.6) {\road{C}};

    \tikzset{shift={(4,0)}}
         
    \draw[roadmarking] (0,0) -- ++(10,0);
    \draw[centreline] (0,0.5) -- ++(10,0);
    \draw[roadmarking] (0,1) -- ++(10,0);
    \node at (5,-0.6) {\road{D}};
    
    \tikzset{shift={(0,3)}}
    \draw[roadmarking] (0,0) -- ++(10,2);
    \draw[centreline] (-.15,0.5) -- ++(10,2);
    \draw[roadmarking] (-.3,1) -- ++(10,2);
    \node at (5,0.25) {\road{E}};
    
    \tikzset{shift={(-.25,4.5)}}
    \draw[roadmarking] (0,0) arc (100:80:30);
    \draw[centreline] (0,0.5) arc (100:80:30);
    \draw[roadmarking] (0,1) arc (100:80:30);
    \node at (5,2.25) {\road{F}};
\end{tikzpicture}
\captionsetup{labelformat=empty}
\caption{Figure 1: Different road geometries of varying diversity.}
\label{fig:roads}
\vspace*{.15cm}
\end{subfigure}
\begin{minipage}[b]{0.48\textwidth}
\begin{example}\label{ex:roadsexample}%
The figure above displays a set of six road geometries, where roads \road{A}, \road{B}, \road{C} and \road{D} will very likely provoke very different driving behaviours, due to the differing curvatures and sharpnesses of turns. Road \road{E}, on the other hand, is a minimally rotated version of \road{D}, which therefore should not change a purely curvature-based \DM when added to the \TS.
\road{F}, on the other hand, is similar but slightly bent. Even though its average curvature is comparable to \road{D} and \road{E}, it will most likely induce different \ac{ads} driving behaviour (``a smooth, long, continuous turn''). An adequately sensitive \DM should therefore reflect the addition of \road{F} to the \TS.%
\end{example}%
\end{minipage}%
\vspace*{-.5cm}
\end{figure}

Likewise, in the \ac{ads} domain many testing approaches aim to generate test suites of driving scenarios that feature various combinations of road structure (\eg road geometry, lane markings), traffic participants (\eg vehicles, pedestrians), and other environmental factors (\eg weather, lighting). Arguably, roads are the most fundamental aspect of a driving scenario, as other aspects will typically be expressed in reference to their geometry (\eg ``overtaking on a straight/curvy road''). Thus, the goal must be to create a \TS of diverse road geometries, to test as many vehicle behaviours as possible. This \TS should ideally cover a range of different straights, bends, curves, and turns of various degrees of sharpness. The testing of \acp{ads} is typically coverage-based, aiming to maximise the number of tested road shapes.\footnote{Note that this research investigates road diversity. Test suite optimisation such as minimality or execution efficiency are orthogonal and considered future work.} An adequate \emph{\DM} should therefore reflect the prioritisation of variety of road geometries. \Cref{ex:roadsexample} illustrates this concept over a set of simple roads. This interpretation of diversity, which is rooted in Software Testing, is different than the one from other disciplines. For instance, in Machine Learning, diversity in training data refers to ``equality of distributions'', which is required to avoid biased datasets. Such a view is also common in some biological and natural settings~\cite{Leinster2012}, \eg animal and plant population measurements.

Existing work proposed several \DMs to express the diversity of roads, including measures based on \emph{\acl{js}}~\cite{gambi2019automatically}, \emph{Iterative Levenshtein distance}~\cite{RiccioFSE2020}, \emph{discrete Fr\'echet distance}~\cite{mosig2005approximately}, and various other road features (\eg \emph{curvature}, \emph{complexity}, and \emph{direction coverage})~\cite{zohdinasab2021deephyperion,sadat2021diverse}.

To the best of our knowledge, the choice of \DM has only been reported in published papers, but never justified. Furthermore, the underlying concept of road diversity measures seems to have never been studied before in the context of \ac{ads} testing. This raises the paramount questions of ``\emph{how much road geometries matter in testing \acp{ads}}?'' and ``\emph{how to measure the diversity of a set of roads?}'' and motivates our research in finding a set of \DMs that can be confidently used as objective measures of road diversity. Our research aims to answer the fundamental questions: \emph{``Which \DMs are well-suited for use in \ac{ads} testing?''} and \emph{``is there any overlap (correlation) between the individual measures?''}. To answer these questions, we analyse the \DMs' properties and check if they are indeed good indicators for vehicle \acl{bd}.

\paragraph{Desired properties of \aclp{dm}}
We believe that, in order to be useful for testing, a \DM should guarantee certain properties which provide guidance towards the achievement of testing goals akin traditional coverage criteria, such as code coverage, which guide automatic test generation~\cite{Fraser2011}. We expect adequate road \DMs to have the following properties.

\begin{itemize}[leftmargin=*]
\item \emph{Monotonicity and growth}~\cite{WeyukerTSE86}: A test suite's diversity should never decrease when adding roads. Monotonicity is easily achieved by classical coverage criteria in which the test requirements to cover are fixed and known in advance. On the contrary, in domains like \ac{ads}, in which test requirements might not be known in advance, ensuring monotonicity is far from trivial. In addition to monotonicity, another aspect that is interesting to assess is \emph{how} a \DM grows as the number of test cases included in a \TS increases. Indeed, a \DM that grows for each individual test is able to discriminate better among roads than a \DM that, although monotonic, increases only for some specific tests. Being able to discriminate better among roads is desirable, as it can possibly lead to test the driving agent in different conditions.
\item \emph{Insensitivity to duplicates}~\cite{weitzman1992diversity}: Additionally, the \DM should be insensitive to duplicates (also known as ``twin property''~\cite{weitzman1992diversity}) in the test suite, \ie the \DM value should neither increase nor decrease\footnote{Absence of decrease is already guaranteed by monotonicity.} if the same road is added twice. Note that this property contradicts \emph{minimality of the test suite}, which is another desired property in testing; however, this is an orthogonal concern that should be treated independently and should not be accounted for by a \DM.
\item \emph{Efficiency}: A \DM should be also efficient to compute, \ie computing a test suite's \DM should take much less time than executing the test suite. While computing the coverage of classical coverage criteria is usually fast (as it consists of checking the coverage of each test requirement individually), computing a road \DM could be expensive, as it may require pairwise comparison of all roads. 
\item \emph{Additivity}: Where computational complexity is inevitable, a minimal property that should be guaranteed is that the \DM computation should be additive, \ie any new road added to the test suite should not require the re-evaluation of the diversity of all the roads, but only of the currently added road.
\end{itemize}

\paragraph{Correlation with \acl{bd}}
In software testing, one of the most desired properties of structural coverage criteria is the ability to expose different behaviours of the SUT~\cite{Fraser2019}. This also holds in our context, where higher \DM values should effectively correlate with different observed behaviours. If a \DM is \emph{insensitive}, it would consider roads that actually trigger different types of \ac{ads} behaviours as similar: if this is the case, by relying on the coverage of the roads alone, some of those behaviours would not be triggered. On the other hand, a road \DM that is \emph{too sensitive} would flag similar roads as different, so wrongly expecting them to trigger different \ac{ads} behaviours. Additionally, such a \DM would lead users to build unnecessary large test suites that do not increase \BD.

\paragraph{Planned study}
With the exploratory study proposed in this report, we aim to fill the lack of road \DM research and investigate their effectiveness in ensuring an \ac{ads}'s behaviour coverage. We therefore analyse the properties of a total of \numDMs \DMs (see \cref{sec:subjects}) that either have been used for measuring road diversity in previous research on testing, or have been used to measure diversity in related domains, such as measuring geometric diversity of lines and curves~\cite{agarwal2014computing}, or diversity in populations~\cite{weitzman1992diversity}. We perform our evaluation empirically on an extensive data set of roads, as mathematical analyses are not always possible (or are highly complex). This empirical study also provides quantitative estimates of effect sizes (\eg value growth and efficiency) using real data, which are difficult to obtain through purely analytical methods.

\section{Background and Related Work}
\label{sec:background}

The notion of diversity has been considered in various forms for the testing of \acp{ads}. Here, we first introduce these different methods in \cref{sec:diversityADStesting}. Then, in \cref{sec:dmRoadGeometry}, we provide an overview of \acp{dm} that can be used to quantify the diversity of road geometries in a test suite.

\subsection{Diversity in \ac{ads} Testing}\label{sec:diversityADStesting}
Several works on \ac{ads} testing aim to generate driving scenarios cost-effectively. Most of these works aim to achieve driving scenarios' diversity to avoid generating too similar scenarios that might expose the same issues multiple times.

Abdessalem et al.~\cite{Abdessalem2018TVC}, Tuncali et al.~\cite{TuncaliHSCC2018}, Zhu et al.~\cite{ZhuTITS2021}, Majumdar et al.~\cite{MajumdarFASE2021}, and Zhong et al.~\cite{Zhong_abs-2109-06126} define diversity in terms of differences in the scenario configuration space. The scenario configuration space includes various parameters such as the initial position and speed of vehicles and pedestrians, the road layout, the placement of scenery elements and obstacles, as well as weather and lighting conditions. 
While Abdessalem et al. and Tuncali et al. did not measure how much scenarios differ, Zhu et al. used Euclidean distance to assess how different the scenarios are in the configuration space. Majumdar et al., instead, defined diversity based on the overall dispersion of the scenario parameters, whereas Zhong et al. proposed to discriminate diverse tests only if they are at a certain distance in the configuration space.

Riccio and Tonella~\cite{RiccioFSE2020} designed an Iterative Levenshtein distance to evaluate test diversity. However, differently from the work mentioned above, they considered only geometric properties of roads, \ie the road shape. Other notable works that considered only road properties as test diversity discriminant are the ones by Zohdinasab et al.~\cite{zohdinasab2021deephyperion}, Nguyen et al.~\cite{DBLP:conf/aitest/NguyenHG21}, Gambi et al.~\cite{SBSTtoolcomp22}\cite{gambi2019automatically}, and Tang et al.~\cite{DBLP:conf/icra/TangZWLSHW21}.
In particular, Zohdinasab et al. computed high-level road features, like smoothness or complexity, and represented the tests into bi-dimensional feature maps such that roads mapped to the same map cells are considered similar. Notably, Zohdinasab et al. also used behavioural features to identify tests that expose similar behaviour of the \ac{ads}. This map representation has been later used for test selection~\cite{DBLP:conf/aitest/NguyenHG21}, and test adequacy assessment~\cite{SBSTtoolcomp22}.
Gambi et al. and Tang et al., instead, discriminated tests based on whether they take place on roads and intersections made of similar road segments. 

Most of the existing approaches that take road geometry into account measure test diversity by aggregating similarity metrics computed over pairs of roads. The next section provides the necessary details on road diversity metrics and aggregation functions.

\begin{table}[!tb]
\centering
\caption{Diversity Measures - Outline}
\label{tab:dm-overview}
{\footnotesize
\resizebox{\columnwidth}{!}{
\begin{tabular}{ccc}
\toprule
\textbf{Distance functions} & \textbf{Aggregation Methods} & \textbf{Direct \DMs} \\
\midrule
Discrete Fr\'{e}chet Distance & Weitzman Aggregation & Test Set Diameter \\
Partial Curve Mapping & Distance Entropy & Convex Hull of Curves \\
Dynamic Time Warping & Summing & \\
Normalised Relative Angle & Averaging & \\
Complexity Vectors & Averaging of Maxima & \\
Iterative Levenshtein Distance &  & \\
Jaccard Similarity Index &  & \\
Area Between Curves &  & \\
Manhattan Distance & & \\
\bottomrule
\end{tabular}}
}
\end{table}

\subsection{Diversity Measures for Road Geometry}\label{sec:dmRoadGeometry}

In the literature, several methods have been described for computing \acp{dm} for roads.
Furthermore, we also draw from other domains such as geometry and population diversity, to obtain potentially viable \DM methods. We categorise these methods as (1) methods that can be computed by \emph{aggregating pairwise distances} between roads, and (2) methods that can be \emph{computed directly} by using the information of each individual road in a test suite. 

\cref{tab:dm-overview} lists the approaches in each category.
Diversity computation methods that belong to the first category are formed as a pair of a distance function and an aggregation method.

In the following, we overview existing distance functions and then explain various methods that can be used to aggregate the distances between road pairs to compute the test suite diversity.

\subsubsection{Pairwise Distance Functions}
To the best of our knowledge, the following functions are commonly used for measuring dissimilarities between roads, (typically represented as curves):

\paragraph{Fr\'echet and discrete Fr\'echet distances} Fr\'echet~\cite{aronov2006frechet} and discrete Fr\'echet~\cite{mosig2005approximately,agarwal2014computing} distance functions are commonly used to measure distance between curves. Intuitively, when two vehicles move along two roads, the \emph{maximum point-to-point distance} between them depends on the speed that they move. Fr\'echet distance corresponds to the shortest of all possible maximum point-to-point distances that can be obtained by varying the speeds. Discrete Fr\'echet distance provides an efficient way of approximating Fr\'echet distance for the case where the curves are described as sequences of straight segments. These distances have been used for characterising road dissimilarity~\cite{roadRepSBTADSsQRS21} and the dissimilarity of the paths vehicles can take~\cite{fan2010frechet,weise2021many}.

\paragraph{Partial curve mapping distance} Partial curve mapping distance between curves representing two roads is defined as the sum of the discrepancy between the segments of two polygons representing normalised versions of the curves~\cite{witowski2012parameter}. 

\paragraph{Dynamic time warping distance} Dynamic time warping was originally proposed for computing dissimilarity between temporal sequences where entries correspond to data obtained at consecutive time instants~\cite{berndt1994using}, but it has also been commonly used for checking dissimilarity between curves~\cite{munich1999continuous,efrat2007curve}.

\paragraph{Relative angle and normalised relative angle distances} Relative angles and normalised relative angles between curves as defined in~\cite{vlachos2004rotation} provide useful methods for computing pairwise distance measures that do not change when the curve representing one of the roads is rotated. This rotation invariance property is especially useful when testing trajectory planners~\cite{werling2010optimal,mcnaughton2011motion,zhu2020trajectory} utilised in \acp{ads}, as they depend on the sharpness of the turns on a road but not the entire orientation of the road (\eg whether the road goes north or east).

\paragraph{Distance based on complexity vectors} In~\cite{sadat2021diverse}, roads were identified as sequences of smaller sections called frames. Then,  curvatures and curvature-derivatives were used  to compute the so-called complexity vectors for each frame.  Given a pair of roads with multiple frames, the distance function of~\cite{sadat2021diverse} finds the maximum of the distances between the complexity-wise closest frames of the roads.

\paragraph{Iterative Levenshtein distance}
Curves representing roads can be described as lists of connected straight segments. The rotational differences between the orientation of consecutive segments form sequences of angles. Iterative Levenshtein distance for a pair of roads is defined in~\cite{RiccioFSE2020} as the Levenshtein \emph{edit distance}~\cite{levenshtein1966binary} between the sequences of angles identified for each of the roads. 

\paragraph{Jaccard similarity index}
\ac{js} for a pair of roads is defined in~\cite{gambi2019automatically} by considering the sets of segments in those roads. \ac{js} takes a value between $0$ and $1$ corresponding to the ratio of the number of common segments of the roads to the total number of segments in the union of segment sets. Thus, the value $1-\mathrm{JS}$ can be used as a distance function quantifying the dissimilarity between a pair of roads. 

\paragraph{Area between curves} The area between curves is a geometric measure that, casually speaking, measures ``how much space fits'' between two curves~\cite{jekel2019}. This measure allows fast computation of the pairwise distance between road curves.

\paragraph{Manhattan distance over feature vectors} Zohdinasab et al.~\cite{zohdinasab2021deephyperion} used Manhattan distance between feature vectors extracted from two roads as a measure of dissimilarity between them.

\subsubsection{Aggregation Methods}
After a distance function is used on each pair of roads in a test suite, an aggregation method can be used for combining the distances to obtain a single value representing the \DM. The properties of such a \DM thus depend not only on the pairwise distance function but also on the method of aggregation. We report below aggregation methods commonly used to measure diversity:

\paragraph{Weitzman aggregation} Weitzman~\cite{weitzman1992diversity} proposed an aggregation method by considering ideal properties of the relationship between diversity measures and underlying pairwise distance functions. The computation of the proposed aggregation method involves set-based recursions, and is known to be slow in general. 

\paragraph{Aggregation through summing} Distances between roads can be aggregated by taking their sum. This method is also used \eg in biology to measure population diversity and corresponds to  \emph{functional attribute diversity}~\cite{chiu2014distance}. 

\paragraph{Aggregation using distance entropy} Distance entropy was proposed by Shi et al.~\cite{shi2015measuring} as a method for quantifying the diversity of a test suite for applications in software testing. When adapted to our problem setting, this method first constructs a weighted relationship graph of roads by using their pairwise distances. The aggregation is then achieved by computing the entropy of the weights in the minimum spanning tree of the relationship graph. 

\paragraph{Aggregation through averaging all and averaging maximum distances} Averaging-based aggregation methods were previously used by~\cite{roadRepSBTADSsQRS21} and \cite{zohdinasab2021deephyperion}. In particular, \cite{roadRepSBTADSsQRS21} considered average of all pairwise distances between roads as an aggregation method in diversity computation. Moreover,  \cite{zohdinasab2021deephyperion} considered a more general setting where pairwise distances are defined for general test cases (not just roads). In the case of roads, the aggregation method of \cite{zohdinasab2021deephyperion} corresponds to calculating the average of the distances from each road to the road that it is most distant to (\ie averaging maximum distances). 

\subsubsection{Direct Computation \acp{dm}}

There are two diversity quantification methods that do not rely on distances between roads.  

\paragraph{Test set diameter} Test set diameter was introduced by Feldt et al.~\cite{feldt2016test} as a method for direct computation of \acp{dm}. It is linked to normalised compression distance for multisets~\cite{calo2020Generating}, which uses Kolmogorov complexity of elements in a set.

\paragraph{Convex hull of curves} Area of the convex hull encompassing all road curves is a \DM that can be computed directly without comparing roads with each other. Previously, convex hulls of curves have been used in~\cite{frenetic2022sbst} for checking if roads fit into a given map.

\subsubsection{Analysis of the Relationship between Different \acp{dm}}\label{sec:mathPropertiesDMs}
The relationship between different \acp{dm} can be characterised through correlation of the diversity values that they assign to test suites. In some special cases, correlation coefficients can be analytically derived. For instance, for the same underlying distance function, \acp{dm} obtained with aggregation through summation and averaging are perfectly correlated (with correlation coefficient 1), since one is a scaled version of the other. Analytical correlation analysis becomes more challenging when \acp{dm} use nonlinear aggregation methods or nonlinear operations in diversity calculations. This point is further discussed in the context of diversity of species in~\cite{debenedictis1973correlations}. There, it is mentioned that while it is possible to show positive correlation between \acp{dm}, the exact value of the correlation coefficient is hard to derive analytically in many cases. In such cases, numerical methods are typically used (see, \eg \cite{mouchet2010functional}). 

Another aspect of correlation analysis of \acp{dm} is that correlation of aggregation-based \acp{dm} depends also on the relationship between the underlying distance functions. We note that for some \acp{dm}, the correlation between distance functions is preserved through aggregation. In particular, the correlation coefficient of two \acp{dm} defined by summing all pairwise distances obtained respectively with two different distance functions is equivalent to the correlation coefficient of the distance functions themselves. However, the correlation coefficient is hard to obtain analytically, since there is no straightforward transformation between distance functions for curves. In~\cite{jekel2019}, some of these distance functions were compared through an empirical study on an optimisation problem.

\section{Research Design}\label{sec:researchDesign}
Given the importance of \DMs in the testing of \acp{ads}, an objective comparison of the methods is of vital interest. Specifically, we are interested in the properties of \DMs, as well as whether and how strongly they are correlated. Additionally, we also would like to investigate the relationship between (road) \DMs and \acf{bd} of the vehicle. To this extent, we select \numDMs \DMs (see \Cref{sec:subjects}) and perform the---to the best of our knowledge---first exploratory study of \DMs for road geometry.

\subsection{Research Questions}

We specifically investigate the following \acp{rq}:
\begin{researchquestions}
\item Which \DMs guarantee beneficial properties such as \emph{monotonicity}, \emph{insensitivity to duplicates}, \emph{efficiency}, and \emph{additivity}? We expect that for a ``good'' \DM, the diversity of a \TS cannot decrease by adding more roads and remains constant when containing road duplicates. Furthermore, a \DM should be efficient to calculate and extend (\ie not require complete recalculation). In this \RQ, we check the \DMs individually for these four properties.
\item Are \DMs correlated among each other (\ie pairwise)? Given that certain \DMs measure similar (\eg geometric) properties, we suspect that some of them might be correlated. This information is of interest, as the calculation of strongly correlated \DMs might be redundant. In this \RQ, we test our assumption by checking which \DMs are correlated, and how strong their correlation is.
\item What is the effect of road length on \DMs? Conceptually, long roads could be seen as compositions of shorter road segments (\eg turns and straights). Thus, certain \DMs might ``average out'' or mask specific distinguishing features of the roads, such as \eg when averaging a road with a short sharp turn after a long straight. On the other hand, geometric \DMs might naturally favour longer roads. For instance, the \emph{area between curves} of a slightly left bent and a slightly right bent road increases with the length of the roads. As this information is of practical interest to developers, who have to be aware of such properties, in this \RQ, we study whether any \DMs are correlated with the length of the road. 
\item Do the \DM values correlate with the simulations' \acl{bd}?  This \RQ specifically investigates the main assumption that road diversity can be used as a proxy for \BD.
\begin{researchquestions}
\item Is there a clear correlation between \DMs and the \BD that is calculated from vehicle observations (\ie acceleration, brake, velocity, steering input, lateral position)?
\item Does low (resp. high) road diversity imply low (resp. high) \BD? While \resq{4.a} investigates general correlation, here we focus specifically on those \TSs with low (resp. high) \DM values and their correlation to \BD. By intuition, one might suspect that while \emph{generally high road diversity does not guarantee high \BD, low road diversity certainly implies low \BD.} This \RQ will specifically evaluate such a correlation.
\item Do \TSs that exert low (resp. high) \BD have similar \DMs? This \RQ can be thought of as ``the inverse'' of \resq{4.b}. We focus specifically on \TSs that yield low (resp. high) \BD values and correlate them with the \DMs of the \TSs that exert them.
\item What is the impact of road length on \BD? Intuitively, we might suspect that \BD is masked on longer roads (similar to the effect described in \resq{3}), and hence, that shorter roads are preferable. In this \RQ we analyse the impact of road length on correlation strength between \DMs and \BD.
\end{researchquestions}
\end{researchquestions}

\subsection{Research Subjects}\label{sec:subjects}

The subject of our research are the \numDMs \aclp{dm} described in \cref{sec:background}. Specifically, we will analyse all {combinations of} the nine {pairwise distance measures} and the five {aggregation methods} (see \Cref{tab:dm-overview}). Moreover, we will also consider the {two direct \DMs}
\begin{enumerate*}[label={}]
\item test set diameter and
\item convex hull of curves.
\end{enumerate*}

We will perform our analyses on all \numDMs \DMs, to get a complete picture of the \DM landscape. Note that some properties for certain \DMs can be deduced \eg due to the nature of their aggregation function (monotonicity is for instance not assured when using \emph{averaging}). Nonetheless, as we additionally aim to investigate the effect size, we will calculate these properties for all \DMs.

Similarly, even though one might intuitively suspect certain \DMs to be strongly correlated (\resq{2}), our goal is to test this hypothesis in a practical setting. The information on which and how strongly the \DMs are correlated is of special interest, since it allows future users to avoid computing redundant \DMs, \ie those that produce very similar (if not equivalent) results.

\subsection{Road Data Set}\label{sec:data}

In the \sbstcomp~\cite{SBSTtoolcomp22}, competitors provide search algorithms that generate non-intersecting two-lane roads which an autonomous driving agent should follow. Each generated road is simulated and provides timestamped observation records of the vehicle's \emph{position}, \emph{velocity}, \emph{steering angle}, \emph{brake} and \emph{throttle} inputs. The roads---cubic interpolations of the Cartesian control points provided by the search algorithms---are handed to a simulator and allow calculation of \emph{length}, \emph{curvature}, \emph{road heading}, \emph{turn count} and aggregations thereof (\eg \emph{max curvature}), etc. Based on this data, we can extract further information such as \emph{relative heading} \wrt the road, \emph{total driven length}, \emph{lateral vehicle position}, as well as \emph{aggregate} values such as minimum, mean, maximum and standard deviation of steering input, as suggested by~\cite{JahangirovaICST2021}.

We use a large data set of \numroads produced in the course of the competition as the basis for our research on road diversity. To generalise our data, we use roads generated randomly, as well as by three road generators, namely \ambiegen, \wogan and \frenetic (see~\cite{SBSTtoolcomp22}), and simulation data provided by executing these roads using two autonomous driving agents (see \cref{sec:driving-agents}).

\paragraph{Data Quality}
A preliminary look into the data showed that both the road data and the simulation information is of high quality.
The created roads have been analysed for self-intersections and maximum curvature by the \sbst pipeline at creation time. Furthermore, we also checked that the simulation data is complete, \ie the observation data for vehicle simulation (position, velocity, acceleration, steering/throttle/brake inputs, etc.) are available for every record, and the recordings have high enough frequency (\eg between 5Hz and 20Hz). Nonetheless, we observed that a limited number of simulations have potentially invalid data. For instance, we discovered a small number of ``faulty'' simulations where the vehicle started driving in the wrong direction. We will thoroughly analyse the data and remove such simulations before running our experiments analysis on the data. To this extent, we will implement automatic ways to identify and remove these individuals from the data set, and we will also manually check the test samples. Furthermore, next to the road and simulation data itself, the \sbst pipeline also produced and recorded some statistical information (driving direction coverage, road curvature measures) for each simulation. We will adapt our scripts to reproduce this data using our own (independent) implementation, thereby increasing confidence in our code and correctness of the existing data.

\subsection{Driving Agents}\label{sec:driving-agents}

To increase the generalisability of our results, we use driving simulations produced by two independent autonomous \acp{da}, \ie \beamng and \davetwo. Both considered agents are widely used in the literature~\cite{gambi2019automatically, RiccioFSE2020, zohdinasab2021deephyperion, SBSTtoolcomp22} and automatically perform the lane keeping task. They are, however, conceptually different (rule-based vs deep learning-based) and, thus, show different driving behaviour.

\beamng is the \ac{da} shipped with the \texttt{BeamNG.tech}\footnote{\url{https://beamng.tech/}} driving simulator. It defines the ego-car's trajectory before the simulation by leveraging a perfect knowledge of the road's geometry. In particular, \beamng plans a trajectory that maximises the car's speed (within pre-defined speed limits), while keeping the vehicle within the right lane as much as possible.

\davetwo exploits a deep learning architecture consisting of three convolutional and five fully-connected layers~\cite{BojarskiTDFFGJM16}. In particular, it learns a direct mapping from the on-board sensor camera input to the steering angle value to be passed to the ego-car's actuators. This means that it does not require previous knowledge of the road geometry. 

In our study, we will perform all our analyses that take driving behaviour into account (\ie \resq{4}) \wrt each specific agent.

\section{Execution Plan}\label{sec:plan}

In the following, we describe our specific research plan and provide the detailed protocol for the experiments for each \RQ in Sections \ref{subsec:rq1}--\ref{subsec:rq4}. Furthermore, shared experimental settings are presented before.

\paragraph{\Aclp{ts}} The analysis of both road and behavioural diversity is based on the calculation of metrics for \TSs. A \TS is a set of randomly selected roads of fixed size.  Our \TSs are sampled with varying sizes of 10, 20, 50 and 100 roads. To generalise our results, we aim to perform each computation on 100 \TS of each size\footnote{Even though we will use powerful computation infrastructure, certain \DMs (\eg those using Weitzman aggregation) are computationally (very) expensive for large \TSs. We will therefore set a maximum computation time and adjust our analysis accordingly.}.

\paragraph{\Acl{bd}} The metric for \ac{bd} aims to express how much of a vehicle's behavioural range is covered within a \TS. In this work, we say that the behaviour of a vehicle on a road is defined by the vehicle's \emph{velocity}, \emph{acceleration}, \emph{braking}, \emph{steering} input and \emph{lateral position} on the road. Each simulation of a \acl{da} on a road produces records of these values in regular intervals (roughly every 0.1 seconds). Through aggregation, we calculate the \emph{mean}, \emph{minimum}, \emph{maximum} and \emph{standard deviation} values of each of the five observations as reported in~\cite{JahangirovaICST2021}, yielding a total of 20 values. For the comparison of \BD of two road simulations, we calculate these 20 values for both roads and compute the (20-dimensional) normalised Euclidean distance. To expand our measurements from pairwise to set-of-roads metrics, we aggregate using the entropy approaches used for output diversity, as described in~\cite{MenendezTSE2022}.

\paragraph{\Acl{da}}
While the results of \resq{1--3} are independent of the specific \acp{da}, for the analysis of \resq{4} we need to take the difference between \acp{da} into account.
Thus, we will separately analyse the correlation of \DMs and \BD for each \ac{da}.

\paragraph{Rotational adjustment}
When thinking of road geometry, we typically only think of the shape of the road itself, but do not take its position and orientation into account. Thus, a perfectly straight road leading north will be seen as equivalent to a straight road of the same length leading east or west. To avoid being misled, we should therefore move all roads to the same starting location and also align them, before calculating the \DMs. Nonetheless, some \acp{da} such as \davetwo (which was trained on image data), take positioning of the sun and ego's own shadow into account for their behaviour. Thus, arguably, roads with the same geometry but different heading should be distinguished when analysing \DMs for \davetwo.

We will therefore duplicate our analyses for the \TSs of \davetwo and report any correlation \wrt rotated as well as non-rotated roads. For the rotation, we will therefore do \emph{data preprocessing} in the form of a \emph{Procrustes analysis}~\cite{dryden2016statistical,roadRepSBTADSsQRS21}, leading to two transformations. First, the roads are relocated so that their starting points match. Second, the roads are rotated around the initial point with the angle of rotation obtained through an optimisation procedure~\cite{dryden2016statistical}. These steps ensure that two roads with identical shapes yield zero distance.

\subsection{RQ1 -- \DM Properties}\label{subsec:rq1}

\paragraph{Monotonicity and growth}
For some \DMs and aggregation methods such as convex hull of curves, Weitzman aggregation and aggregation through summation, monotonicity can be mathematically proved. For these \DMs and aggregation methods, we can simply report the theoretical results known from the literature. For other \DMs and aggregation methods such as test set diameter and aggregation using distance entropy, instead, assessing monotonicity is more challenging as no theoretical results are known from the literature. For this latter category, we  perform an empirical evaluation. 

Moreover, in case a measure is monotonic, we are also interested in assessing ``how and how much'' it grows; specifically, we are interested in assessing to what extent each new test increases \TS  diversity. Indeed, measures for which each novel (non-duplicated) test increases the \DM value are better at discriminating among different tests than \DMs for which only a few, highly diverse tests lead to a noticeable change of the \DM value; better discrimination among tests is desirable, as it can lead to exercise different \acs{ads} behaviours (think of adding road \road{F} in \cref{ex:roadsexample}). Note that in the case of \acp{dm} via aggregation, diversity largely depends on the underlying complex distance functions, and, as a result, a precise mathematical assessment of the growth is hard to achieve. This is the case even for measures for which we can analytically check monotonicity (\eg Weitzman aggregation~\cite{weitzman1992diversity}). Therefore, to quantify the growth, we perform an empirical evaluation. Namely, we evaluate the relative \DM growth when adding new roads to the \TS. Intuitively, smaller \TSs should, on average, have a larger growth in diversity, due to the larger probability of adding a highly diverse road. Nonetheless, even large test suites' \DM values should reflect the addition of individuals. We proceed as follows:
\begin{enumerate*}
\item For each of the \TSs (grouped by size), we compute the \DMs.
\item We add $n$ new roads to each \TS, where $n$ corresponds to $n \in [10\%, 20\%, 50\%, 100\%]$ of the \TS's size and re-compute the \DMs.
\item We then analyse the difference before and after extension, to report typical statistical metrics (mean, min, max, standard deviation) for each \DM and \TS size.
\end{enumerate*}

\paragraph{Insensitivity to duplicates}
Similarly to monotonicity, for some \DMs, this property can be determined directly from the \DM definition whereas for other \DMs an empirical approach is more suitable. In particular, mathematical properties of the distance functions can be used to show that \acp{dm} obtained through Weitzman aggregation and aggregation through summation are insensitive to duplication, but \acp{dm} obtained through averaging-based aggregation methods do not possess the insensitivity property, as duplication can decrease diversity for those \acp{dm}. Moreover, in case insensitivity to duplicates is not guaranteed, it is not possible to mathematically establish to what extent a \DM is sensitive to duplication; therefore, we plan to assess this effect by means of an empirical study. We proceed as follows:
\begin{enumerate*}
\item We calculate the \TSs' \DMs.
\item We then randomly duplicate $n$ roads in each \TSs, where $n$ corresponds to $n \in [10\%, 20\%]$ of the \TS's size and re-compute the \DMs.
\item We then check if any \DM's value changed and analyse the difference in magnitude before and after the extension, using standard descriptive statistics (mean, min, max, standard deviation).
\end{enumerate*}

\paragraph{Efficiency}
Although asymptotic efficiency of a DM can be obtained analytically, it cannot predict concrete results on real data. Therefore, we conduct an empirical assessment to precisely compare the different \DMs and provide an initial evaluation of the \DMs' computational efficiency. We proceed as follows. For each \TS, we record the time it takes to compute each \DM. Due to the total of 100 \TSs of each size, we then confidently analyse and compare the \DM computation times by size, listing their aggregated information (mean, median, standard deviation).

\paragraph{Additivity} Additivity measures the time it takes to re-calculate a \TS's \DM after being extended. As for efficiency, only an asymptotic analysis of additivity is possible starting from the \DMs definition; in order to have a more realistic comparison of the different \DMs, we conduct an empirical study as follows.
\begin{enumerate*}
\item First we calculate each \TSs' \DMs. 
\item We add $n$ new roads to each \TS, where $n$ corresponds to $n \in [1 \text{~road}, 2 \text{~roads}, 10\%, 20\%]$ (Note the addition of 1 and 2 roads).
\item We re-calculate the \DMs and record the time, before
\item we analyse the results.
\end{enumerate*}

\subsection{RQ2 -- Pairwise Correlation of \DMs}\label{sec:rq2Assessment}
Using the \DMs calculated for each \TS, we perform correlation analyses on the totality of the data and grouped by \TS size. 
As discussed in \cref{sec:mathPropertiesDMs}, analytical derivation of correlation coefficients for certain pairs of \acp{dm} is prohibitively complex. For this reason, we perform an empirical evaluation of correlations. We select the correlation test to use depending on the analysed data. If the data is normally distributed, we use the Pearson's correlation test. Instead, if the data is not normally distributed, we use the Spearman's correlation test. Both tests produce as output a number in $[-1, 1]$, where $0$ means that there is no correlation, while $-1$ and $1$ indicate perfect negative and positive correlation. Intermediate values indicate different degrees of correlation, and can be interpreted using existing classifications, such as slight, low, moderate, high, and very high correlation~\cite{cohen2013statistical}. When analysing correlation results between \DMs, we report the corresponding classes and use these to draw conclusions. For instance, if two \DMs are not or only weakly correlated, it could mean that they measure different characteristics of road geometries and that their use in combination might be more beneficial than using either \DM individually. Instead, if two \DMs are strongly correlated, it may hint at redundancy, and the properties of \resq{1} should be used to choose a better-suited one.

\subsection{RQ3 -- Correlation of \DMs and Road Length}
Here we analyse the effect of road length on the individual \DMs.
\begin{enumerate*}
\item We sample random \TSs based on the length of the roads.
As we are specifically interested in short (resp. long) roads, we have to adjust our \TS sampling algorithm to assemble random \TSs from the shortest (resp. longest) quantile of the roads.
\item We then calculate \DMs for all \TSs and perform an analysis using the same statistical tests described in \cref{sec:rq2Assessment}, \ie
\item we calculate \DMs for test suites and 
\item perform correlation analysis (as in \resq{2}) between the \DM values and the (\TSs' average) road length.
\end{enumerate*}

The results of this analysis allow determining whether there is an effect of the road length on the diversity. Combined with the results of \resq{4.d}, it helps users in selecting appropriate road lengths for \ac{ads} testing.

\subsection{RQ4 -- Correlation of \DMs and \BD}\label{subsec:rq4}
We analyse whether there is a correlation between the individual \DMs and the \BD. Unfortunately, due to the complexity of \acp{ads}, it is typically and generally not possible to define a reliable mathematical model of an \ac{ads}'s behaviour from which a correlation can be analytically deduced. Hence, we empirically study such a correlation and test the sensitivity of \ac{ads} behaviour to road diversity. This \RQ aims to discover those \DMs that are actually linked to \BD, thus helping testers avoid using \DMs that focus on irrelevant road features.

\paragraph{\resq{4.a}}
First, we analyse the general correlation between road diversity (\ie \DMs) and \BD. Remember that as \BD is different for the individual \acp{da}, we perform the analyses separately. Specifically, we proceed as follows:
\begin{enumerate*}
\item For each \TS, we calculate the \DMs and 
\item then, for each \TS we calculate the \BD (as described above).
\item Finally, we perform correlation analysis between these values.
\end{enumerate*}

\paragraph{\resq{4.b}} We analyse whether there is a specific relation between \TSs with low (resp. high) diversity and the \BD they expose. Therefore, for each \DM we
\begin{enumerate*}
\item first select the \TSs that have low (resp. high) diversity according to the specific \DM and 
\item analyse the variance of the \TSs' \BD through correlation.
\end{enumerate*}

\paragraph{\resq{4.c}} Intuitively, we perform the ``inverse analysis'' of \resq{4.b}. Specifically, we 
\begin{enumerate*}
\item select test suites that yielded low (resp. high) \BD values in \resq{4.a}, and 
\item evaluate whether the \TSs that exposed them have similarly low (resp. high) \DM values.
\end{enumerate*}

\paragraph{\resq{4.d}} We study the effect of road length on the correlation between \DMs and \BD, to evaluate if for \TSs with short roads, there is a stronger correlation between \DMs and \BD. Hence, we proceed as follows:
\begin{enumerate*}
\item Using the similar-in-length \TSs created in \resq{3}, we calculate for each \TS the values of the \DMs and the \BD (as in \resq{4.a})
\item We then perform a correlation analysis of each \DM and \BD pair.
\item Finally, we compare the yielded correlation coefficients and check if the correlation of \DMs and \BD of \TSs with short (resp. long) roads is higher (resp. lower) than the one of ``normal'' \TSs.
\end{enumerate*}

\section{Discussion}

\subsection{Contributions and Impact}
Our study will help in identifying a catalogue of \DMs that are effective in ensuring behavioural diversity of the considered agents. Such \DMs can be used for multiple tasks in the \ac{ads} testing process. Effective \DMs can drive the generation of test suites that exercise diverse behaviours of the \ac{sut}, \eg as a fitness function of a search-based test input generation algorithm. Another possible usage of the proposed \DMs is the selection of a reduced number of diverse tests from an existing test suite, thus reducing the (high) cost of \ac{ads} testing. 

\paragraph{Code availability}
Evidently, the results of our analysis are specific to the implementation, the evaluation environment and data set used. In the spirit of open science, we make our analysis code publicly available, so other users and researchers can 
\begin{enumerate*}[label={(\alph*)}]
\item reproduce the results of our study,
\item validate the algorithms and implementation of our \DMs, and 
\item extend this study by applying our correlation analyses to other diversity measures, behavioural metrics and data.
\end{enumerate*}

\subsection{Threats to Validity}\label{sec:threats}

The validity of our study may be affected by different types of threats~\cite{Wohlin2012} that we discuss in the following.

\paragraph{Construct validity}
A construct validity threat could be that the approaches we use to investigate our research questions are not appropriate. 
If this is the case, we may draw wrong conclusions regarding the investigated RQs; for example, we may observe correlation where there is none. To mitigate this threat, we carefully design the experiments to answer the RQs. To assess monotonicity, we will explicitly check that the diversity does not decrease when increasing the number of considered roads. For monotonic measures, we will check how they grow. We will proceed similarly for ``insensitivity to duplicates''. In order to check correlation, we will use either the Pearson's correlation test or the Spearman's correlation test depending on whether the data is normally distributed or not. Moreover, we will also consider the strength of the correlation when reporting the results. With regards to measures of \BD, a threat could be that the ones we selected are not representative of the behavioural diversity in the \acp{ads} that we use, or that they do not generalise. This could threaten the conclusions we draw in \resq{4}. To address this threat, instead of coming up with our own \BD measures, we looked at literature~\cite{JahangirovaICST2021} to identify and select \BD measures that have been used and studied before.

\paragraph{Internal validity}
An internal validity threat could be that the observed correlation (if any) between road diversity and behaviour diversity is due to other confounding factors. For example, the behaviour of a driving agent depends not only on the road, but also on other elements of the driving scenarios such as road participants, regulatory elements, weather conditions, etc. In complex scenarios in which all these elements are present, it would very difficult (if possible at all) to isolate and measure the influence of the road structure on the driving behaviour. Therefore, there is the risk that we wrongly observe the correlation because the behaviour diversity is triggered by the different elements of the driving scenarios and not the road structure. To mitigate this threat, we use scenarios composed of only one road, without any other elements. The correlation of behaviour diversity with other traffic elements should be part of independent studies and we leave it as future work.

Other confounding factors could be the test suite size and the length of the road. Regarding test suite size, as for some metrics (\ie those guaranteeing monotonicity) more tests in general lead to higher values, there is the risk of observing a positive correlation that, however, is only due to the test suite size. To mitigate this threat, we analyse the data by considering different test suite sizes separately, as described in \cref{sec:plan}. Regarding the length of the road, instead, we mitigate this threat by assessing the influence of road length on \DMs (\resq{3}) and \BD (\resq{4.d}).

\paragraph{External validity}
A threat of this type is that the results of our study could be not generalisable to other types of roads, and other driving agents. To mitigate this threat, we sample from a very large set of roads created by different road generators, that can possibly generate roads of different shapes. Moreover, to assess the influence of road length on the diversity and correlation with behaviour diversity, we will artificially shorten roads to generate roads of different lengths. Different agents could react differently to different types of roads and so lead to different behaviour diversity values, and so different correlation with road diversity. To mitigate this threat, we consider two very different types of driving agent: \beamng, a rule-based driving agent exploiting the complete knowledge of the road geometry, and \davetwo, a deep learning-based agent designed at NVIDIA~\cite{BojarskiTDFFGJM16}.

\section{Conclusions}
As complete testing of a modern \acp{ads} is impossible, current testing practices aim to expose a \ac{sut} to a suite of diverse scenarios, based on the assumption that high scenario diversity leads to a wide \acl{bd} of the \ac{sut}. In this report, we propose to conduct the first exploratory study to test this assumption. Due to the complexity of \ac{ads} scenarios, we focus our study on the arguably most fundamental part of a scenario description, namely its road geometry. In the past, a variety of road \DMs has been applied in literature, but to the best of our knowledge, this use has only been reported, but never truly justified nor studied deeply.

The present report describes the \aclp{rq} and execution plan for an exploratory study on \numDMs road geometry \DMs that are either currently used in \ac{ads} testing or might be potentially beneficial to it. Specifically, our \aclp{rq} analyse \DMs' properties (\resq{1}), their pairwise correlation (\resq{2}), the relationship between \DMs and road length (\resq{3}) and test the assumption that road diversity induces \acl{bd} (\resq{4}).

Our empirical analyses target a large data set of \numroads individual road geometries and matching simulation data from two distinct driving agents. Based on this data, we will identify a catalogue of \DMs that are effective in ensuring behavioural diversity for \ac{ads} agents and help developers effectively test their systems.

\begin{acks}
S. Klikovits, E. Castellano, and P. Arcaini are supported by ERATO HASUO Metamathematics for Systems Design Project (No. JPMJER1603), JST, Funding Reference number: 10.13039/501100009024 ERATO. S. Klikovits is also supported by \grantsponsor{20K23334}{Japan Society for the Promotion of Science (JSPS)}{} Grant-in-Aid for Research Activity Start-up No~\grantnum{20K23334}{20K23334}. P. Arcaini is also supported by Engineerable AI Techniques for Practical Applications of High-Quality Machine Learning-based Systems Project (Grant Number JPMJMI20B8), JST-Mirai. A. Cetinkaya is supported by JSPS Grant-in-Aid for Early-Career Scientists No.~20K14771. A. Gambi was partially supported by the DFG project STUNT (DFG Grant Agreement n. FR 2955/4-1) and is also supported by the EU Project FLEXCRASH (Grant agreement ID: 101069674).
\end{acks}

\bibliographystyle{ACM-Reference-Format}
\bibliography{bibliography}

\end{document}